\documentclass[twocolumn,showpacs,aps,dvipdmx]{revtex4.1}
\usepackage{graphicx}
\usepackage{bm}
\usepackage{amsmath}
\usepackage{amssymb}
\usepackage{color}
\usepackage{epstopdf}
\def\l{\langle}
\def\r{\rangle}

\begin{document}
\title{
Comparison of diluted antiferromagnetic Ising models \\
on frustrated lattices in a magnetic field
} 
\author{Konstantin Soldatov$^{1,2}$}
\email{soldatov_ks@students.dvfu.ru}
\author{Alexey Peretyatko$^{1}$}
\email{peretiatko.aa@dvfu.ru}
\author{Konstantin Nefedev$^{1,2}$}
\email{nefedev.kv@dvfu.ru}
\author{Yutaka Okabe$^{3}$}
\email{okabe@phys.se.tmu.ac.jp}
\affiliation{
$^1$School of Natural Sciences, Far Eastern Federal University, Vladivostok, 
Russian Federation \\
$^2$Institute of Applied Mathematics, Far Eastern Branch, 
Russian Academy of Science, Vladivostok, Russian Federation \\
$^3$Department of Physics, Tokyo Metropolitan University, Hachioji, 
Tokyo 192-0397, Japan \\
}

\date{\today}

\begin{abstract}
We study diluted antiferromagnetic Ising models 
on triangular and kagome lattices in a magnetic field, 
using the replica-exchange Monte Carlo method. 
We observe {\it seven} and {\it five} plateaus in the magnetization curve 
of the diluted antiferromagnetic Ising model on the triangular 
and kagome lattices, respectively, when a magnetic field is applied. 
These observations contrast with the two plateaus observed 
in the pure model.  The origin of multiple plateaus is investigated 
by considering the spin configuration of triangles in the diluted models. 
We compare these results with those of a diluted antiferromagnetic 
Ising model on the three-dimensional pyrochlore lattice 
in a magnetic field pointing in the [111] direction, 
sometimes referred to as the "kagome-ice" problem. 
We discuss the similarity and dissimilarity of the magnetization curves 
of the "kagome-ice" state and the two-dimensional kagome lattice.
\end{abstract}


\maketitle

\section{Introduction}

Geometrical frustration, a phenomenon associated with structural 
topology, has recently generated considerable interest. 
Spin-ice materials, such as pyrochlores Dy${_2}$Ti${_2}$O${_7}$ 
and Ho${_2}$Ti${_2}$O${_7}$, have attracted particular attention 
\cite{Harris,Ramirez,Bramwell}. 
In these materials, the magnetic ions (Dy$^{3+}$ or Ho$^{3+}$) occupy 
sites in a pyrochlore lattice formed of corner-sharing tetrahedra. 
Their local crystal-field environment causes their magnetic moments 
to orient along the directions connecting the centers of 
adjacent tetrahedra at low temperatures. 
In the low-temperature spin-ice state, the magnetic
moments are highly constrained locally and obey the so-called
``ice rules", whereby two spins point in and two spins point out of each
tetrahedron of the pyrochlore lattice.
This two-in two-out spin configuration is analogous to 
hydrogen atoms in water ice \cite{Pauling}.

The effects of magnetic field on the spin-ice materials, 
in particular the formation of magnetization plateaus, 
have been studied both theoretically \cite{Harris98,Moessner,Isakov} 
and experimentally \cite{Matsuhira02,Hiroi,Sakakibara,Higashinaka,Fukazawa}. 
The effects of dilution on frustration were studied by Ke {\it et al.} 
\cite{Ke} in spin-ice materials.  In those studies, magnetic ions 
Dy or Ho were replaced by nonmagnetic Y ions. 
The experiments revealed a non-monotonicity in the zero-point entropy, 
considered as a function of the dilution concentration. 
Further studies of the dilution-related effects 
have also been reported \cite{Lin,Scharffe,Shevchenko}.

Quite recently, Peretyatko, Nefedev, and Okabe \cite{Peretyatko}
studied the effects of a magnetic field on diluted spin-ice materials 
in order to elucidate the interplay of dilution and magnetic field. 
They observed {\it five} plateaus in the magnetization curve of the diluted 
nearest-neighbor spin-ice model on the pyrochlore lattice 
when a magnetic field was applied in the [111] direction. 
This effect contrasts with the case of a pure (i.e.~nondiluted) model, 
which displays two plateaus.  The origin of the {\it five} plateaus 
was investigated by considering the spin configuration of 
two corner-sharing tetrahedra in a diluted model. 

One question of interest is whether the existence of multiple steps in the 
magnetization curve is specific to the diluted model, 
applied to the three-dimensional (3D) pyrochlore lattice. 
Another question is what happens in the magnetization curve of 
diluted antiferromagnetic (AFM) Ising models of two-dimensional (2D) 
frustrated lattices.  
The purpose of the present paper is to study the diluted AFM Ising model 
in the presence of a magnetic field, in the case of 
triangular and kagome lattices. 
The calculation for the 2D models is the same 
as for the pyrochlore lattice. 
We performed a similar analysis of the origin of multiple plateaus 
for the case of the pyrochlore lattice, considering 
the spin configuration in a triangle instead of a tetrahedron. 

The pyrochlore lattice can be regarded as an alternating sequence 
of kagome and triangular layers that become effectively decoupled 
by a magnetic field oriented along the [111] direction. 
In this model, the Zeeman energy is expressed as 
$- \bm{h} \cdot \bm{d}_{\kappa(i)}$, where $\bm{d}_{\kappa(i)}$ 
is a unit vector of each spin pointing one of the four corners 
of tetrahedron from the center \cite{Isakov,Peretyatko}. 
When the magnetic field is applied in the [111] direction, 
say, $\bm{h} = h \bm{d}_0$, $\bm{h} \cdot \bm{d}_{\kappa(i)}$ 
becomes $h$ for apical spins with $\bm{d}_{\kappa(i)} = \bm{d}_0$, 
but $-(1/3)h$ for other spins. 
The spins in the triangular layers, apical spins, are fixed 
when the magnetic field 
is applied in this direction. The behavior of the spins 
in the kagome layers is therefore of significant interest, 
and is sometimes referred to as the "kagome-ice" problem. 
It is instructive to study the similarity and dissimilarity 
between the magnetization curves of the "kagome-ice" state and 
the 2D kagome lattice.

In this study, we applied the diluted AFM 
Ising model to the triangular and kagome lattices 
in a magnetic field.
We compare the results with those of the pyrochlore lattice.  
Our simulations were based on the replica-exchange Monte Carlo 
method \cite{Hukushima} to avoid the system becoming trapped 
in local-minimum configurations.

The paper is organized as follows: 
Section II describes the model and the method. 
The results are then presented and discussed in Sec.~III. 
Section IV concludes with a summary and discussion.  

\section{Model and Simulation Method}

\begin{figure}[t]
\begin{center}
\includegraphics[width=4.8cm]{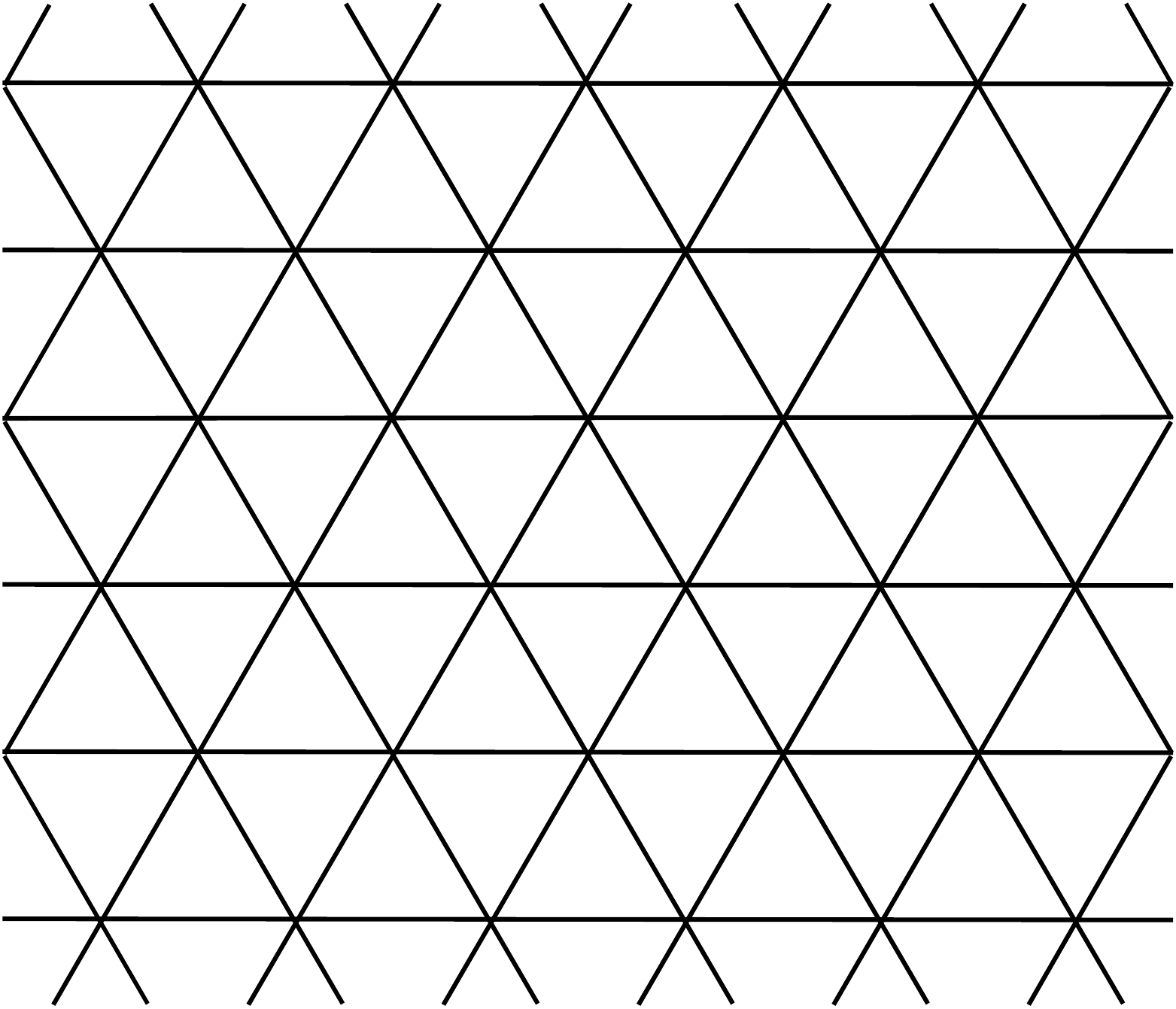}
\hspace{0.3cm}
\includegraphics[width=3.2cm]{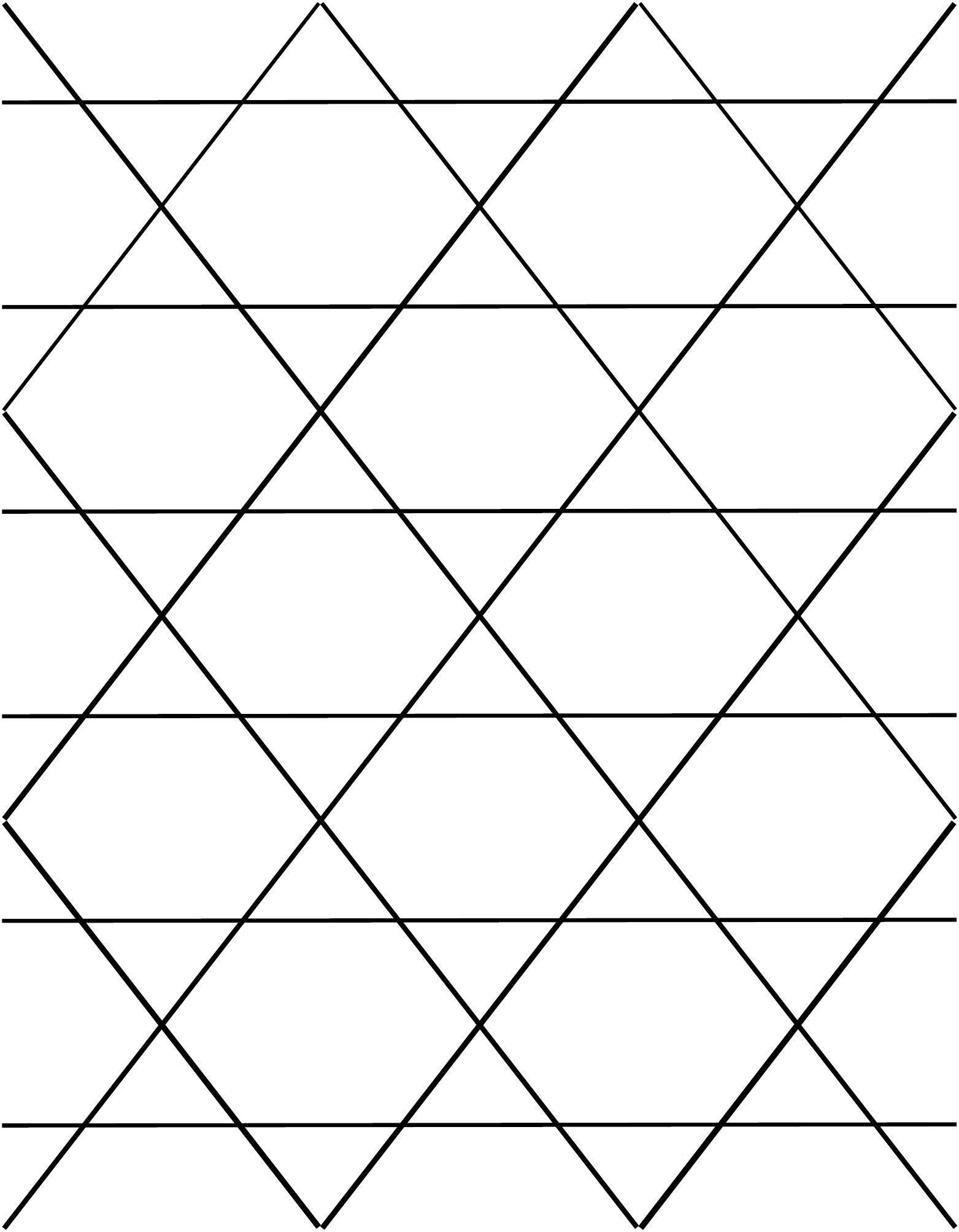}
\caption{
Illustrations of the triangular lattice (left) 
and the kagome lattice (right).
}
\label{fig:tri}
\end{center}
\end{figure}

We studied the AFM Ising model in a magnetic field, 
applied to 2D triangular and kagome lattices. 
The corresponding Hamiltonian is given by
\begin{equation}
 H = J \sum_{\l i,j \r} \sigma_i \sigma_j
   -  h \sum_i \sigma_i, 
\end{equation}
where $J (>0)$ represents the AFM coupling, $h$ is the magnetic field, 
$\sigma_i$ are the Ising spins ($\sigma_i = \pm 1$), and 
$\l i,j \r$ denotes a nearest-neighbor pair. 
Triangular and kagome lattices are illustrated 
in Fig.~\ref{fig:tri}, for convenience.
We here focus on the effects of site dilution on the frustrated 
AFM Ising models.  The Hamiltonian then becomes
\begin{equation}
 H = J \sum_{\l i,j \r} c_i c_j \sigma_i \sigma_j
   - h \sum_i c_i \sigma_i, 
\end{equation}
where $c_i$ are the quenched variables ($c_i = 1 \ {\rm or} \ 0$) 
and the concentration of vacancies is denoted by $x$. 

We use the Monte Carlo simulation in the present study. 
The replica-exchange Monte Carlo method \cite{Hukushima} 
is particularly effective in avoiding local-minimum traps. 
In a previous study, which considered the AFM Ising model 
on the pyrochlore lattice in a magnetic field applied 
in the [111] direction \cite{Peretyatko}, 
the replica-exchange method was successfully applied. 
We here employ the same simulation method. 
We perform the replica exchange of both temperature and 
magnetic field. 
After each Monte Carlo step of single spin updates, 
replica exchanges are carried out. 
When we exchange the inverse temperatures $\beta_1$ and $\beta_2$ 
of two replicas, 
we use a transition probability based on the relative 
Boltzmann weight, such as 
\begin{eqnarray*}
  &~&\exp[-(\beta_1 E_2+\beta_2 E_1)+(\beta_1 E_1+\beta_2 E_2)] \\
   &~&\hspace{1cm} =\exp[(\beta_1-\beta_2)(E_1-E_2)]. 
\end{eqnarray*}
Here, $E_1$ and $E_2$ are the total energies of two replicas.
When we exchange the magnetic fields $h_1$ and $h_2$, 
the relative Boltzmann weight becomes 
\begin{eqnarray*}
  &~&\exp[-\beta (h_1 E^{\rm{(Z)}}_2+ h_2 E^{\rm{(Z)}}_1)
          +\beta (h_1 E^{\rm{(Z)}}_1+ h_2 E^{\rm{(Z)}}_2)] \\
   &~&\hspace{1cm} =\exp[\beta (h_1-h_2)(E^{\rm{(Z)}}_1-E^{\rm{(Z)}}_2)],
\end{eqnarray*}
where $hE^{\rm{(Z)}}$ is the Zeeman energy part.
The calculation may involve many replicas 
of several temperatures and several magnetic fields. 
In the present simulation, we treated 486 replicas of 
81 magnetic fields and 6 temperatures simultaneously 
for the triangular lattice, and 336 replicas of 
56 magnetic fields and 6 temperatures for the kagome lattice. 

The simulation of the AFM Ising model on the triangular lattice 
considered systems of size $L \times L$ with periodic boundary conditions, 
with $L = 48$ ($N = 2304$) and $L = 96$ ($N = 9216$). 
For the kagome lattice of size $L \times (3/2)L$, we used $L = 48$ 
($N = 3456$) and $L = 96$ ($N = 13824$). 
The dilution concentrations $x$ were 
$x$= 0.0 (pure), 0.1, 0.2, 0.4, 0.6, and 0.8. 
We discarded the first 5,000 Monte Carlo Steps (MCSs) 
to avoid the effects of initial configurations, 
and used the next 50,000 MCSs to generate the measurements.
Statistical errors were estimated by calculating averages 
over 20 samples for each size and each $x$ value.

\section{Results}
\subsection{Triangular lattice}

\begin{figure}
\begin{center}
\includegraphics[width=8.0cm]{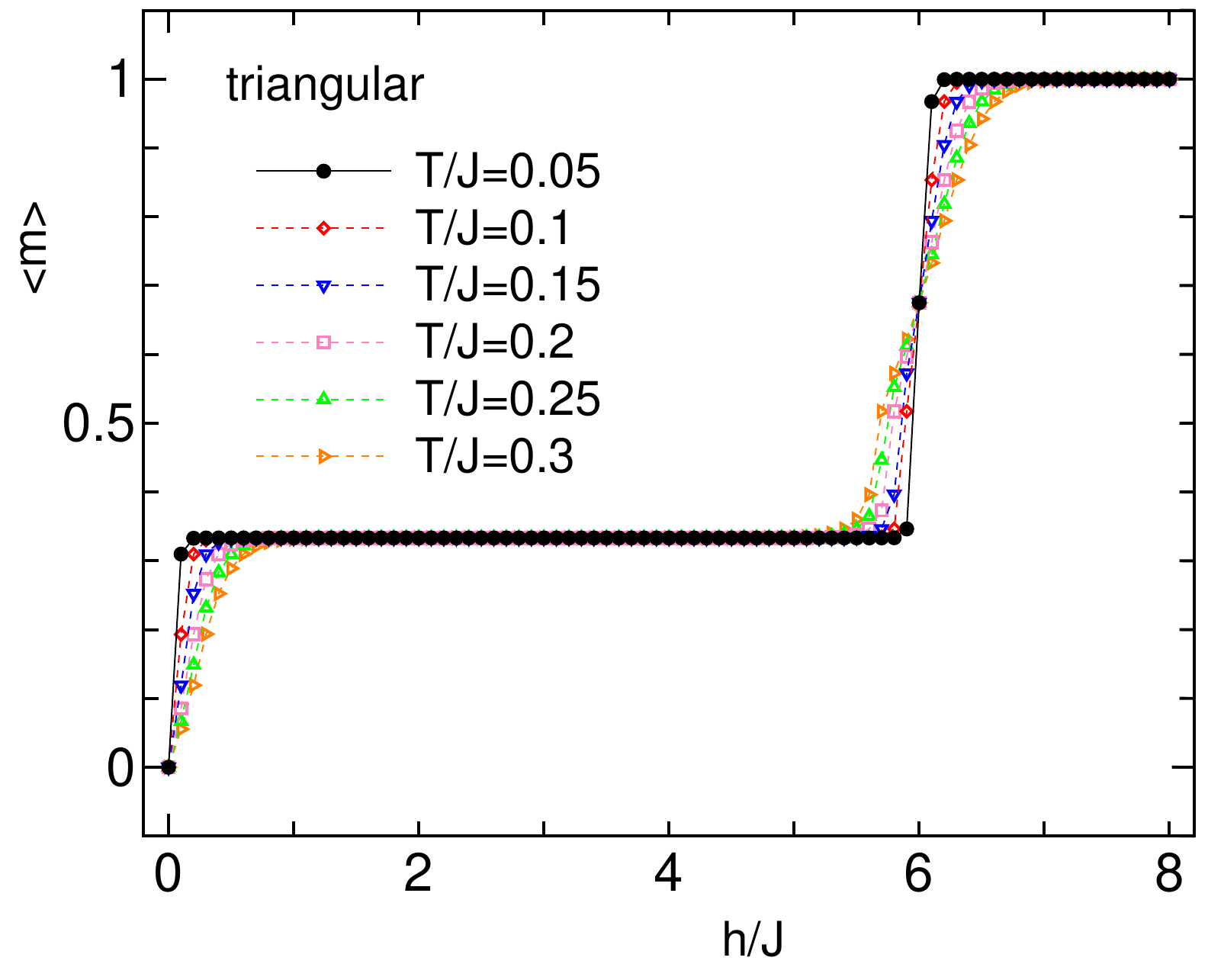}
\caption{
Magnetization curve for the AFM Ising model 
on the triangular lattice. 
The system size is $L=96$ ($N=9216$), 
and the temperatures are $T/J$ = 0.05, 0.1, 0.15, 0.2, 0.25 and 0.3. 
}
\label{fig:tri_pure_mag}
\end{center}
\end{figure}

We first consider the results for the triangular lattice. 
The averaged values of the magnetization $m=M/N$ with $M = \sum_i \sigma_i$
for the pure model ($x=0.0$) are plotted 
in Fig.~\ref{fig:tri_pure_mag} as a function 
of the applied field $h$ expressed in units of $J$.  
The system size is $L=96$ ($N=9216$). 
The temperatures are $T/J$ = 0.05, 0.1, 0.15, 0.2, 0.25, and 0.3. 
Averages were computed over 20 samples 
with different random-number sequences. 
The statistical errors are smaller than the size of the marks.
The size dependence is small for large enough sizes, such as 
$L=48$ and $L=96$. 
We see an $m=1/3$ plateau for $h/J<6$, 
and the jump becomes smoother with increasing temperature.

\begin{figure}
\begin{center}
\includegraphics[width=8.0cm]{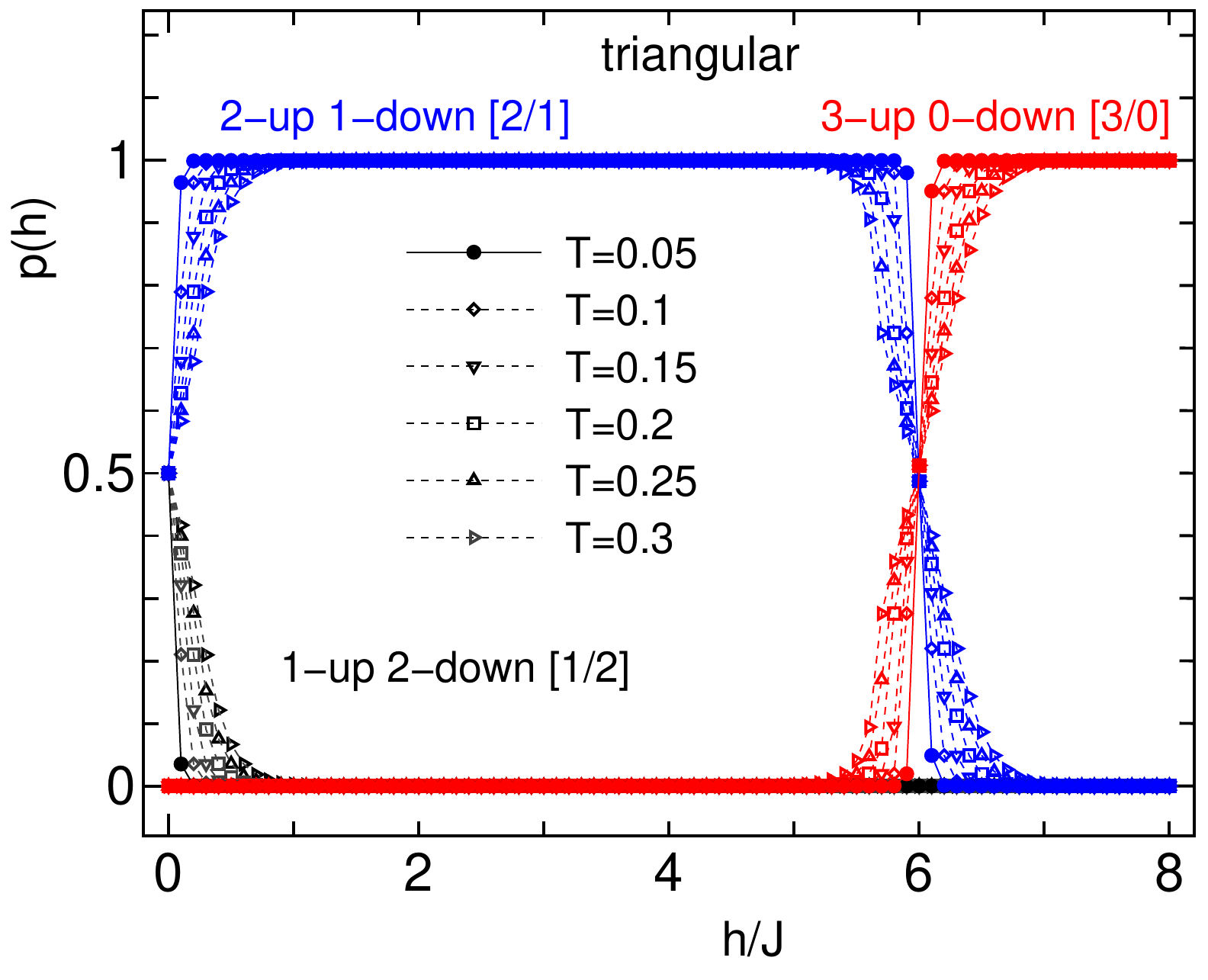}
\caption{
Proportions of the types of spin configurations in the triangle 
for the AFM Ising model on the triangular lattice 
in a magnetic field. 
The system size is $L=96$ ($N=9216$), 
and the temperatures are $T/J$ = 0.05, 0.1, 0.15, 0.2, 0.25, and 0.3. 
}
\label{fig:tri_pure_config}
\end{center}
\end{figure}

Figure~\ref{fig:tri_pure_config} displays the proportions of 
the types of spin configurations within the triangle, 
for the AFM Ising model on the triangular lattice 
in the magnetic field. 
There are 18432 triangles for $L=96$, and the proportions of 
the types of spin configurations were measured for 50,000 MCSs. 
The type of a given spin configurations was characterized in terms of 
the numbers of up ($+1$) spins and down ($-1$) spins 
within the triangle.  
There is a clear transition from the 2-up 1-down 
configuration to the 3-up 0-down configuration 
at $h/J=6$. This transition becomes smoother 
with increasing temperature.

\begin{figure}
\begin{center}
\includegraphics[width=8.0cm]{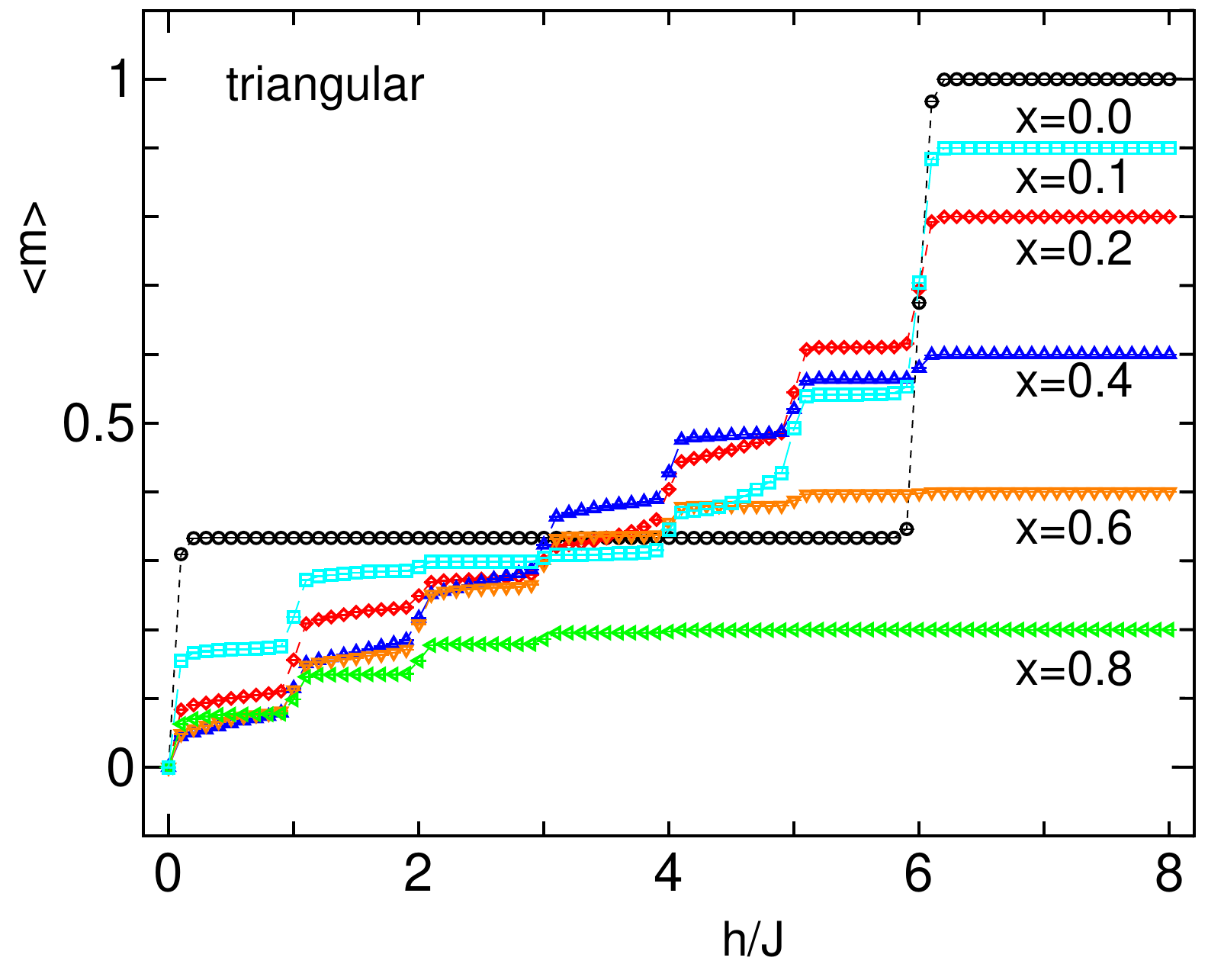}
\caption{
Magnetization curve for the diluted AFM Ising model 
on the triangular lattice. The system size is $L=96$ ($N=9216$), 
and the temperature is $T/J=0.05$. 
The dilution concentrations ($x$) are 0.0, 0.1, 0.2, 0.4, 0.6 and 0.8.
}
\label{fig:tri_dilution_mag}
\end{center}
\end{figure}

The magnetization curve for the diluted AFM Ising model 
on the triangular lattice is plotted in Fig.~\ref{fig:tri_dilution_mag}. 
The system size is $L=96$ ($N=9216$), 
and the temperature is $T/J=0.05$. 
The dilution concentrations ($x$) are 0.0, 0.1, 0.2, 0.4, 0.6 and 0.8.
Averages were calculated over 20 random samples. 
The error bars in the figure are smaller than the size of marks. 
The statistical errors, obtained by averaging over 20 random samples, 
become very small when the system size is as large as $L=96$ ($N=9216$).

We observe seven plateaus in the magnetization curve of diluted 
systems; 
for $h/J<1$, $1<h/J<2$, $2<h/J<3$, $3<h/J<4$, $4<h/J<5$, 
$5<h/J<6$, and $h/J>6$. 
In contrast, the pure case shows only two plateaus, 
located on either side of $h/J = 6$. 
The results shown correspond to $T/J=0.05$. 
At higher temperatures, the magnetic step between the plateaus 
becomes smoother.
The magnetization $m$ saturates at $(1-x)$.

\begin{figure}[t]
\begin{center}
\includegraphics[width=8.4cm]{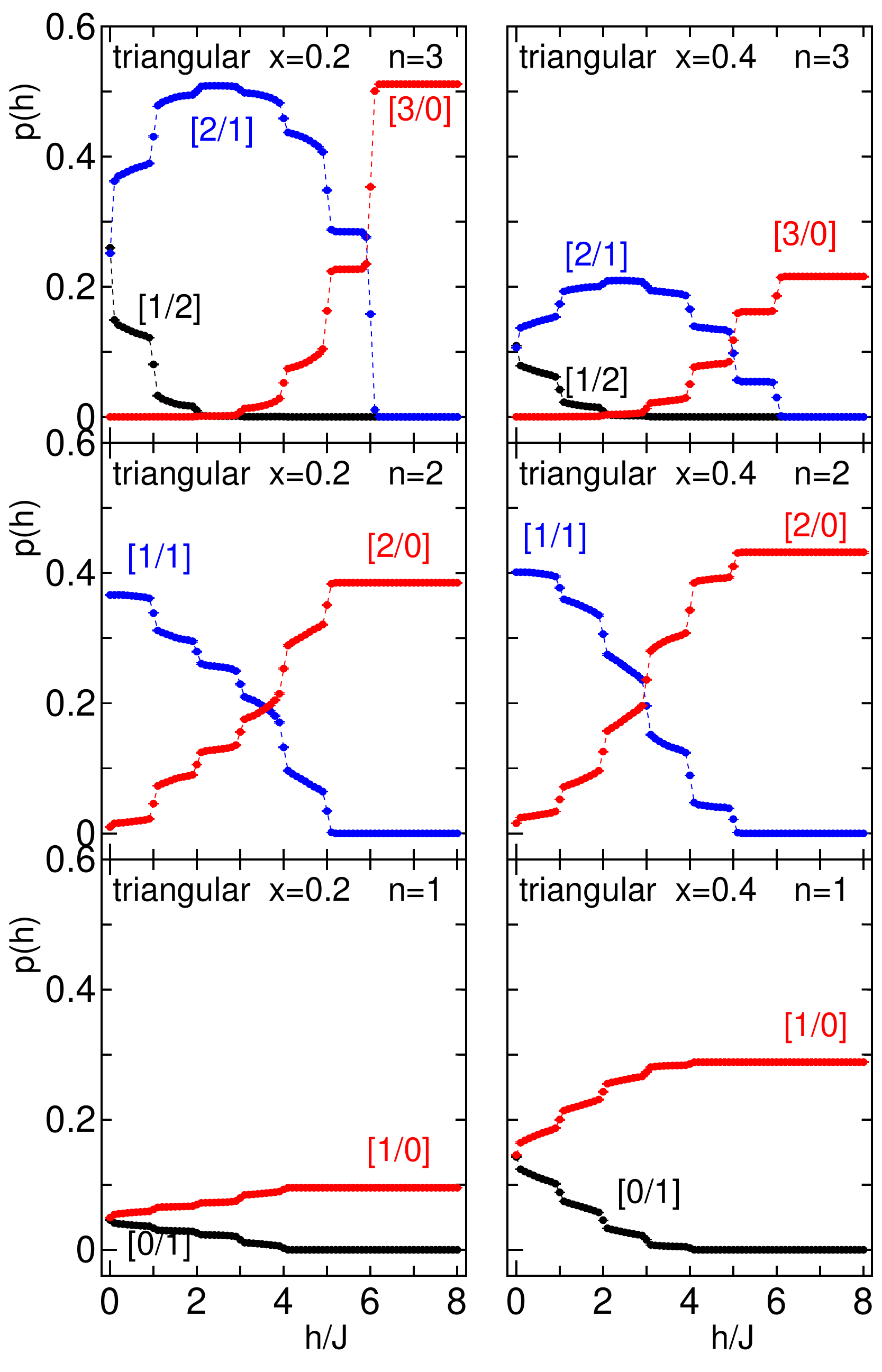}
\caption{
Proportions of the types of spin configurations in the triangle 
for the diluted AFM Ising model on the triangular lattice 
in a magnetic field. 
The system size is $L=96$ ($N=9216$), 
and the temperature is $T/J$ = 0.05. 
The dilution concentrations ($x$) are 0.2 (left) and 0.4 (right).
The number of spins $n$ in the triangle is 3, 2, and 1 
for the top, middle, and bottom panel, respectively. 
}
\label{fig:tri_dilution_config}
\end{center}
\end{figure}

In an earlier study of the diluted triangular-lattice AFM Ising model 
in a magnetic field, Yao \cite{Yao} obtained the magnetization curve 
for the diluted system of weak dilution regions using the Wang-Landau method. 
\u{Z}ukovi\u{c} {\it et al.} \cite{Zukovic} 
used the effective-field theory to study the diluted AFM Ising model. 
Multiple plateaus were reported in both studies. 
Effects of thermal fluctuations were also discussed \cite{Zukovic,Borovsky}.

In the case of the pure model, the spin configuration of the triangle 
changes from the 2-up 1-down configuration to the 3-up 0-down configuration 
when the magnetic field is applied. 
On the contrary, the spin configuration becomes more complex 
in diluted systems.  
The magnetic-field dependence of the spin configuration 
is plotted in Fig.~\ref{fig:tri_dilution_config} 
for a system size $L=96$ ($N=9216$), 
and the temperature $T/J$ = 0.05. 
The dilution concentrations ($x$) are 0.2 and 0.4.
Averaging was performed over 20 random systems.
There are 18432 triangles for $L=96$, 
and the number of spins in a triangle, $n$, becomes 
$n$ = 3, 2, 1, or 0 for the diluted systems.

The top panel of Fig.~\ref{fig:tri_dilution_config} 
corresponds to $n=3$. 
There are $n = 3$ spins for approximately 51\% of all triangles 
when $x = 0.2$, and approximately 22\% when $x=0.4$. 
The change from the 2-up 1-down configuration, [2/1], to 
the 3-up 0-down configuration, [3/0], is observed 
at $h/J=6$ as in the pure case.  
However, the proportions of [2/1] and [3/0] also change 
at $h/J$ =1, 2, 3, 4, and 5. 
For the low-$h$ region, the 1-up 2-down configuration, $[1/2]$, 
remains. 

The middle panel of Fig.~\ref{fig:tri_dilution_config} 
corresponds to $n=2$. 
One spin is deleted from the triangle. 
The partial transition from the 1-up 1-down configuration, [1/1], to 
the 2-up 0-down configuration, [2/0], is observed. 

The bottom panel of Fig.~\ref{fig:tri_dilution_config} 
corresponds to $n=1$. 
Here, two spins are deleted from the triangle. 
At $h/J=0$, there are equal proportions of up spins and down spins, 
whereas the proportion of up spins 
increases as the magnetic field $h$ increases. 

We showed the data of the proportions of the types of 
spin configurations in the triangle for $x=0.2$ and $x=0.4$ 
in Fig.~\ref{fig:tri_dilution_config}. 
We see the $x$ dependence.  The situation is essentially the same, 
although the proportions of configurations with smaller $n$ values 
increase as $x$ increases.  For strong dilution, 
such as $x=0.8$, many free spins appear, 
which does not produce a magnetization plateau; 
the magnetization jump decreases for larger $x$.

\begin{table}
\caption{
The local energy of the spin configuration in the triangle 
for the triangular lattice. 
The spin numbers $n$ are 3, 2, and 1. 
An edge is shared with two triangles, 
and a corner is shared with six triangles.
}
\label{tri_configuration}
\begin{center}
\begin{tabular}{lllll}
\hline
\hline
config. \quad & $n$ spins \quad & up \quad\quad & down \quad & energy \\
\hline
$[3/0]$  & \ 3 & \ 3 & \ 0 & \ $3(J/2)-3(h/6)$ \\
$[2/1]$  & \   & \ 2 & \ 1 & \ $-(J/2)- (h/6)$ \\
$[1/2]$  & \   & \ 1 & \ 2 & \ $-(J/2)+ (h/6)$ \\
$[2/0]$  & \ 2 & \ 2 & \ 0 & \ $ (J/2)-2(h/6)$ \\
$[1/1]$  & \   & \ 1 & \ 1 & \ $-(J/2)$ \\
$[1/0]$  & \ 1 & \ 1 & \ 0 & \ $-(h/6)$ \\
$[0/1]$  & \   & \ 0 & \ 1 & \ $+(h/6)$ \\
\hline
\end{tabular}
\end{center}
\end{table}

We follow a similar procedure as for the case of the pyrochlore 
lattice in the [111] magnetic field \cite{Peretyatko} 
to elucidate the origin of the seven plateaus in the magnetization curve. 
We investigate the local energy of the spin configuration 
in a triangle for $n$ = 3, 2, and 1, the results of which are shown 
in Table \ref{tri_configuration}. The local energy for each configuration 
is given in the last column. An edge, which is related 
to the exchange energy, is shared by two neighboring triangles, 
and a corner, which is related to the Zeeman energy, 
is shared by six triangles. 

\begin{figure}[t]
\begin{center}
\includegraphics[width=8.4cm]{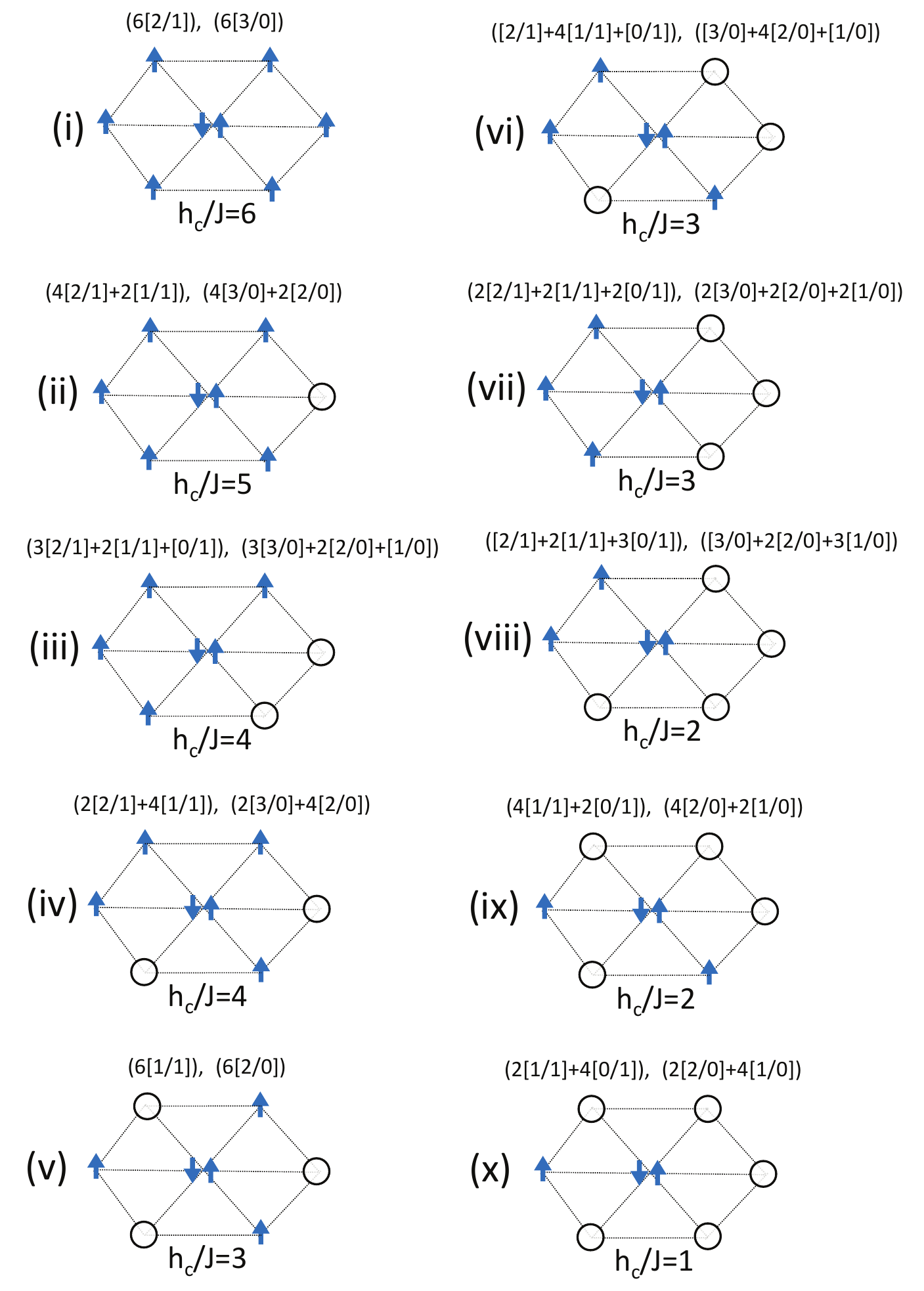}
\caption{
Schematic illustration of spin flip for the triangular lattice.
The up and down spins are denoted by the arrow, 
whereas the deleted spins are denoted by the empty circle.
The crossover values $h_c$ are given there.
}
\label{fig:tri_dilution_flip}
\end{center}
\end{figure}

We consider the energy of six corner-sharing triangles. 
Figure~\ref{fig:tri_dilution_flip} illustrates the flipping process 
schematically. 
The deleted spins are denoted by empty circle.
The case where all the six triangles have $n=3$ is shown in panel (i). 
When the corner-sharing spin is turned from "down" to "up", 
the configuration changes from [2/1] to [3/0]. 
From Table \ref{tri_configuration}, the crossover magnetic field 
is calculated by
\begin{eqnarray*}
 &~& -(J/2)-(h/6) =3(J/2)-3(h/6). 
\end{eqnarray*}
We then obtain $h_c/J=6$.
If one spin is deleted as in panel (ii), there are four triangles 
with $n=3$ and two triangles with $n=2$. 
When the corner-sharing spin is turned from "down" to "up", 
the configuration changes from (4[2/1]+2[1/1]) to 
(4[3/0]+2[2/0]). 
From Table \ref{tri_configuration}, the crossover magnetic field 
is calculated by
\begin{eqnarray*}
 &~& 4(-(J/2)-(h/6))+2(-(J/2)) \\
 &~&\hspace{1cm} =4(3(J/2)-3(h/6))+2((J/2)-2(h/6)). 
\end{eqnarray*}
We then obtain $h_c/J=5$. 
The deletion of two spins makes us consider two cases: 
firstly, the change from (3[2/1]+2[1/1]+[0/1]) to (3[3/0]+2[2/0]+[1/0]), 
as shown in panel (iii), and secondly, 
the change from (2[2/1]+4[1/1]) to (2[3/0]+4[2/0]), as shown in panel (iv). 
In both cases, the crossover magnetic field is $h_c/J=4$.
The cases where three spins are deleted are shown in panels 
(v), (vi), and (vii), 
whereas those where four spins are deleted are shown in panels (viii) and (ix). 
Panel (x) corresponds to the deletion of five spins. 
The crossover values $h_c$, given in Fig.~\ref{fig:tri_dilution_flip},
can be obtained in the same manner. 

This investigation clearly accounts for the change in the 
configurations displayed in Fig.~\ref{fig:tri_dilution_config}, 
which clarifies the origin of the seven magnetization plateaus 
in Fig.~\ref{fig:tri_dilution_mag}. 
The magnetization step at $h/J=6$ comes from the situation 
shown in panel (i) of Fig.~\ref{fig:tri_dilution_flip}. 
When one increases the dilution concentration $x$, 
the possibility of such spin configurations becomes small. 
Thus, for strong dilution cases, the magnetization steps 
at $h/J=6$ and also at $h/J=5$ become very small.

\subsection{Kagome lattice}

\begin{figure}
\begin{center}
\includegraphics[width=8.0cm]{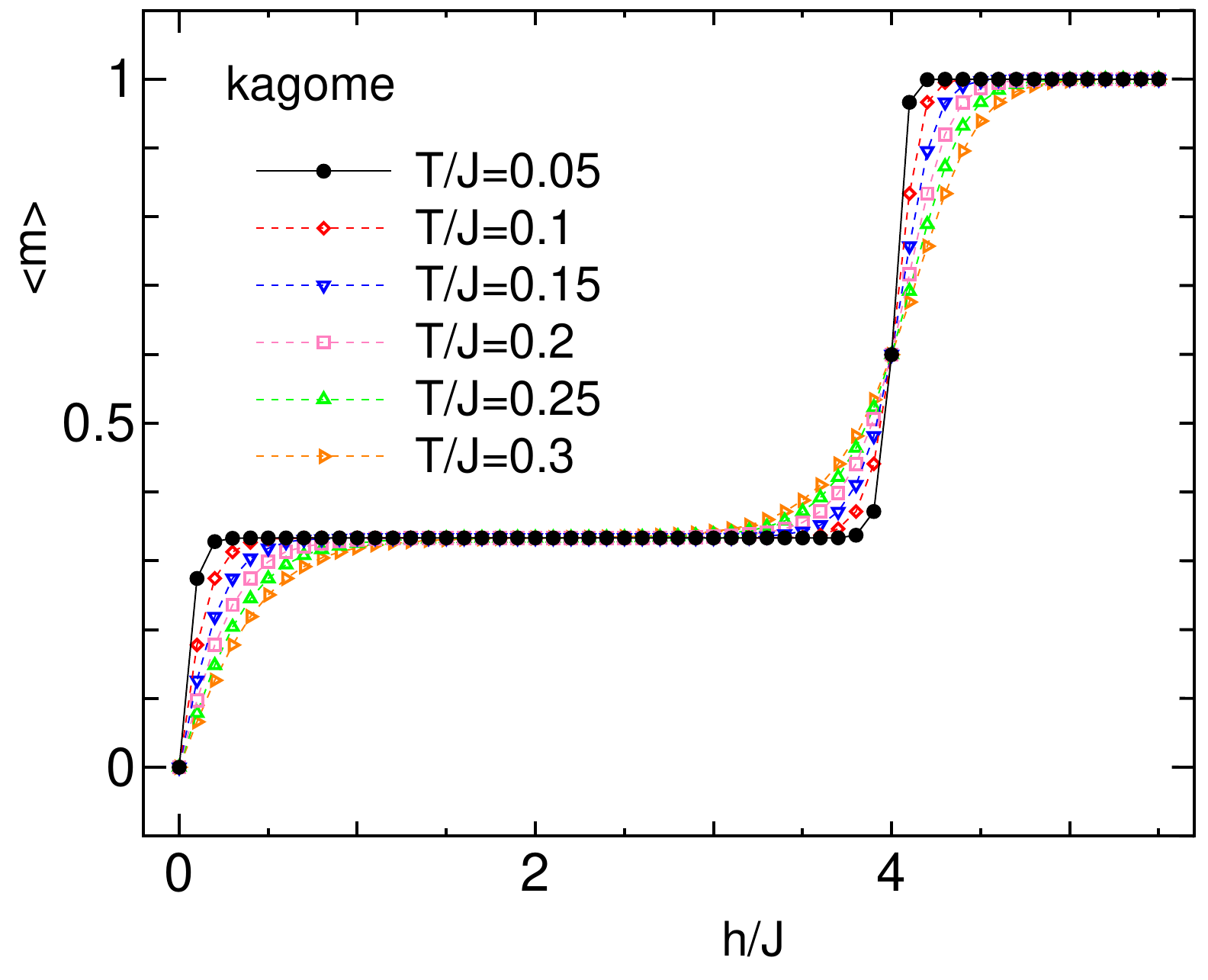}
\caption{
Magnetization curve for the AFM Ising model 
on the kagome lattice. 
The system size is $L=96$ ($N=13824$), 
and the temperatures are $T/J$ = 0.05, 0.1, 0.15, 0.2, 0.25 and 0.3. 
}
\label{fig:kag_pure_mag}
\end{center}
\end{figure}

We next consider the kagome lattice. 
The averaged values of the magnetization $m=M/N$ for the pure model 
($x=0.0$) are plotted as a function of the applied field $h$ 
in Fig.~\ref{fig:kag_pure_mag}. 
The system size is here $L=96$ ($N=13824$) and 
the temperatures are $T/J$ = 0.05, 0.1, 0.15, 0.2, 0.25, and 0.3. 
Averaging was performed over 20 samples 
with different random-number sequences. 
The plateau corresponding to $m=1/3$ appears for $h/J<4$, 
but the jump becomes smoother with increasing temperature.

\begin{figure}
\begin{center}
\includegraphics[width=8.0cm]{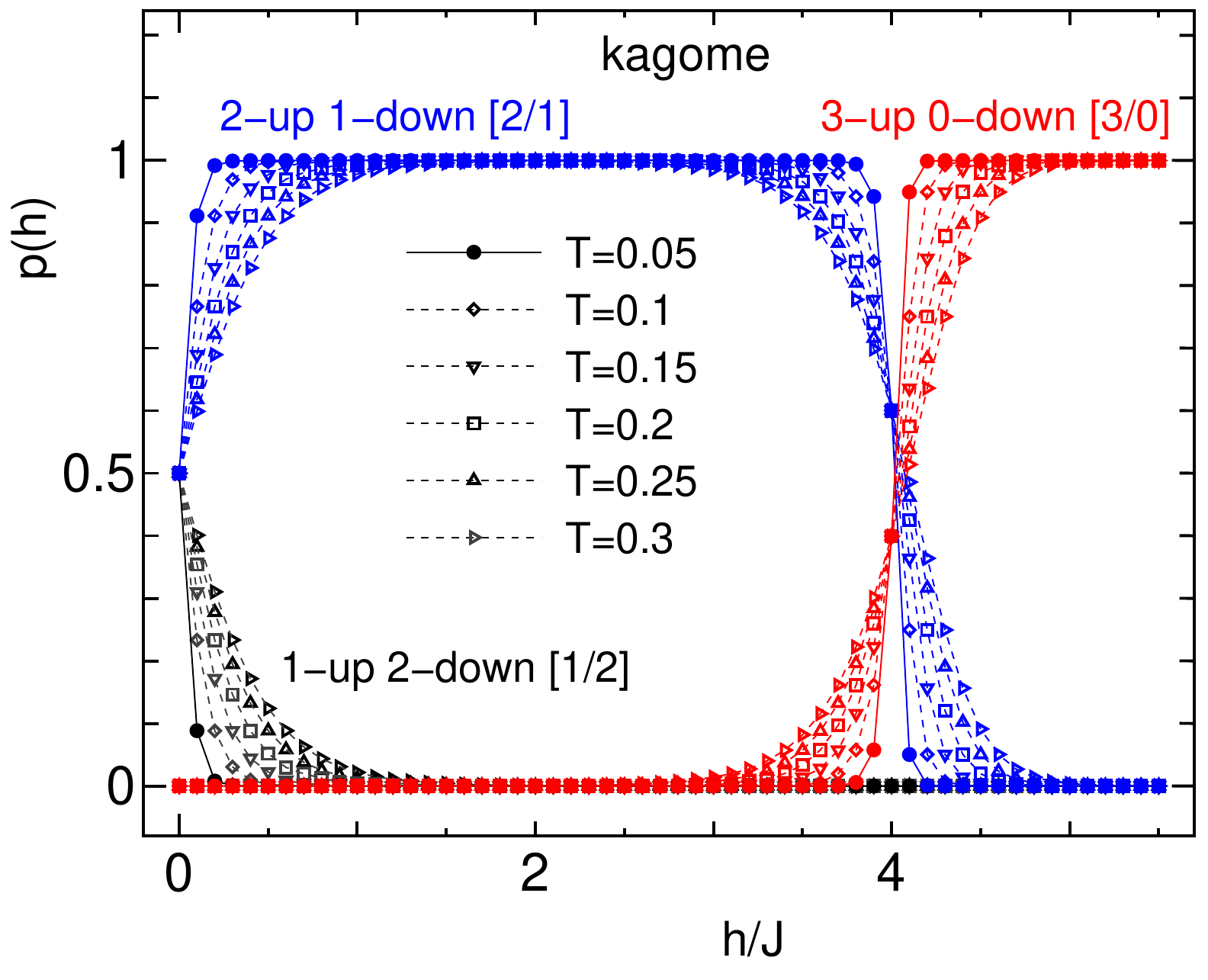}
\caption{
Proportions of the types of spin configurations in the triangle 
for the AFM Ising model on the kagome lattice 
in a magnetic field. 
The system size is $L=96$ ($N=13824$), 
and the temperatures are $T/J$ = 0.05, 0.1, 0.15, 0.2, 0.25, and 0.3. 
}
\label{fig:kag_pure_config}
\end{center}
\end{figure}

Figure~\ref{fig:kag_pure_config} shows the proportions of the 
different types of spin configurations in a triangle 
for the AFM Ising model on the kagome lattice 
in a magnetic field. 
There are 9216 triangles for $L=96$, and the proportions of 
the different types of spin configurations were measured for 50,000 MCSs. 
The type of spin configurations in a triangle is classified 
in the same way as for the triangular lattice. 
There is a clear transition from the 2-up 1-down 
configuration to the 3-up 0-down configuration 
at $h/J=4$. The transition becomes smoother 
when increasing temperature.

\begin{figure}
\begin{center}
\includegraphics[width=8.0cm]{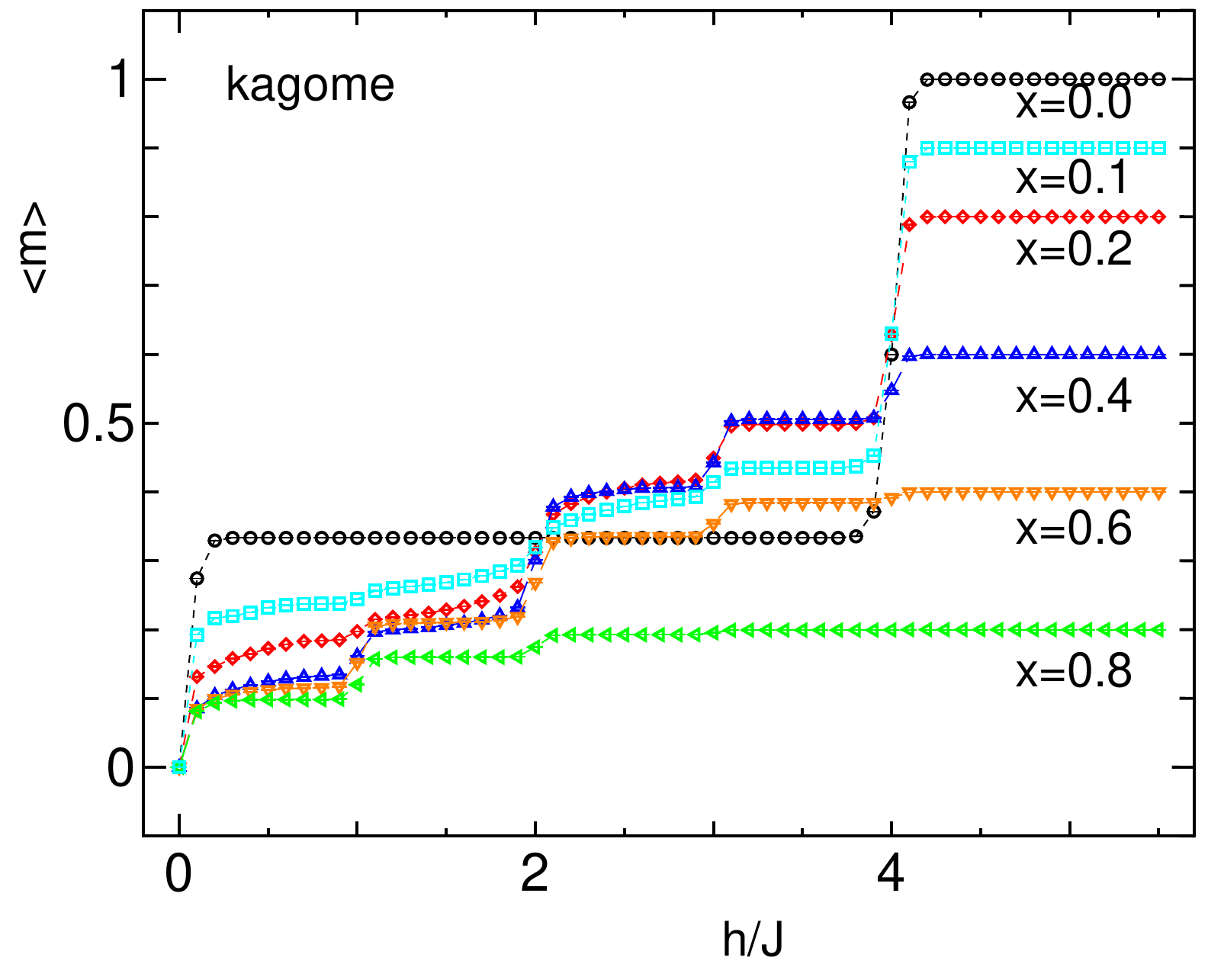}
\caption{
Magnetization curve for the diluted AFM Ising model 
on the kagome lattice. The system size is $L=96$ ($N=13824$), 
and the temperature is $T/J=0.05$. 
The dilution concentrations ($x$) are 0.0, 0.1, 0.2, 0.4, 0.6 and 0.8.
}
\label{fig:kag_dilution_mag}
\end{center}
\end{figure}

The magnetization curve for the diluted AFM Ising model 
on the kagome lattice is plotted in Fig.~\ref{fig:kag_dilution_mag}. 
The system size is $L=96$ ($N=13824$), 
and the temperature is $T/J=0.05$. 
The dilution concentrations ($x$) are 0.0, 0.1, 0.2, 0.4, 0.6 and 0.8.

Figure~\ref{fig:kag_dilution_mag} shows five plateaus 
in the magnetization curves, in the ranges 
$h/J<1$, $1<h/J<2$, $2<h/J<3$, $3<h/J<4$, and $h/J>4$. 
This situation contrasts with the pure case, where 
only two plateaus are observed, on either side of $h/J=4$.
The saturated value of the magnetization $m$ 
is $(1-x)$.

\begin{figure}[t]
\begin{center}
\includegraphics[width=8.4cm]{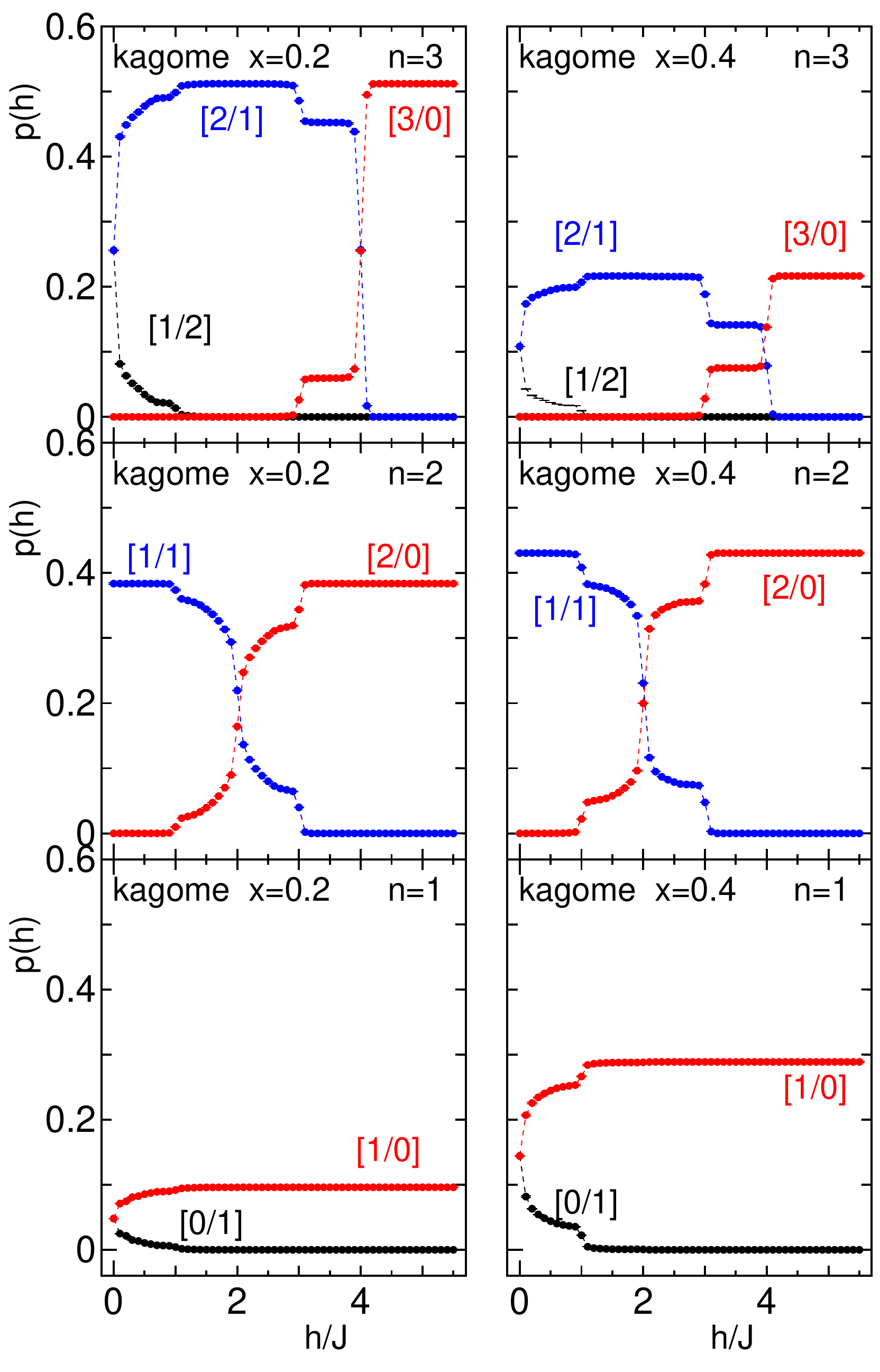}
\caption{
Proportions of the types of spin configurations in the triangle 
for the diluted AFM Ising model on the kagome lattice 
in a magnetic field. 
The system size is $L=96$ ($N=13824$), 
and the temperature is $T/J$ = 0.05. 
The dilution concentrations ($x$) are 0.2 (left) and 0.4 (right).
The number of spins $n$ in the triangle is 3, 2, and 1 
for the top, middle, and bottom panel, respectively. 
}
\label{fig:kag_dilution_config}
\end{center}
\end{figure}

The magnetic-field dependence of the spin configuration 
for the diluted AFM model on the kagome lattice 
is plotted in Fig.~\ref{fig:kag_dilution_config}.
The system size is $L=96$ ($N=13824$), 
and the temperature is $T/J$ = 0.05. 
The dilution concentrations ($x$) are 0.2 and 0.4.
There are 9216 triangles for $L=96$, 
and the number of spins in the triangle, $n$, becomes 
$n$= 3, 2, 1, or 0 for the diluted systems.

The top panel of Fig.~\ref{fig:kag_dilution_config} shows 
the result for $n=3$, which corresponds to approximately 
51\% of all the triangles for $x=0.2$ and approximately 
22\% for $x=0.4$ again. 
A transition from the 2-up 1-down configuration, [2/1], to 
the 3-up 0-down configuration, [3/0], is observed 
at $h/J=4$, as in the pure case.  
However, the proportions of the [2/1] and [3/0] configurations 
also change at $h/J$ =1, 2, and 3. 

The middle panel of Fig.~\ref{fig:kag_dilution_config} corresponds 
to $n=2$. One spin is deleted from the triangle. 
The partial transition is observed from the 1-up 1-down configuration, 
[1/1], to the 2-up 0-down configuration, [2/0]. 

The bottom panel corresponds to $n=1$,  
where two spins are deleted from the triangle. 
At $h/J=0$, there are equal proportions of up and down spins,  
whereas the proportion of up spins 
increases with increasing $h$. 

We showed the data of the proportions of the types of 
spin configurations in the triangle for $x=0.2$ and $x=0.4$. 
The situation is essentially the same, 
although the proportions of smaller $n$ in the triangle 
increases when $x$ become larger.  

\begin{table}
\caption{
The local energy of the spin configuration in the triangle 
for the kagome lattice. 
The spin numbers $n$ are 3, 2, and 1. 
An edge is not shared with other triangles, 
and a corner is shared with two triangles.
}
\label{kag_configuration}
\begin{center}
\begin{tabular}{lllll}
\hline
\hline
config. \quad & $n$ spins \quad & up \quad\quad & down \quad & energy \\
\hline
$[3/0]$  & \ 3 & \ 3 & \ 0 & \ $3J-3(h/2)$ \\
$[2/1]$  & \   & \ 2 & \ 1 & \ $-J- (h/2)$ \\
$[1/2]$  & \   & \ 1 & \ 2 & \ $-J+ (h/2)$ \\
$[2/0]$  & \ 2 & \ 2 & \ 0 & \ $ J-2(h/2)$ \\
$[1/1]$  & \   & \ 1 & \ 1 & \ $-J$ \\
$[1/0]$  & \ 1 & \ 1 & \ 0 & \ $-(h/2)$ \\
$[0/1]$  & \   & \ 0 & \ 1 & \ $+(h/2)$ \\
\hline
\end{tabular}
\end{center}
\end{table}

We use the same analysis as above to elucidate the origin of 
the multiple plateaus in the magnetization curve.  
The local energies of the spin configuration of the triangle 
for $n$= 3, 2, and 1 are 
listed in Table \ref{kag_configuration}. 
An edge, which is related to the exchange energy, is not shared with 
other triangles, 
and a corner, which is related to the Zeeman energy, 
is shared between two triangles.

\begin{figure}[t]
\begin{center}
\includegraphics[width=8.4cm]{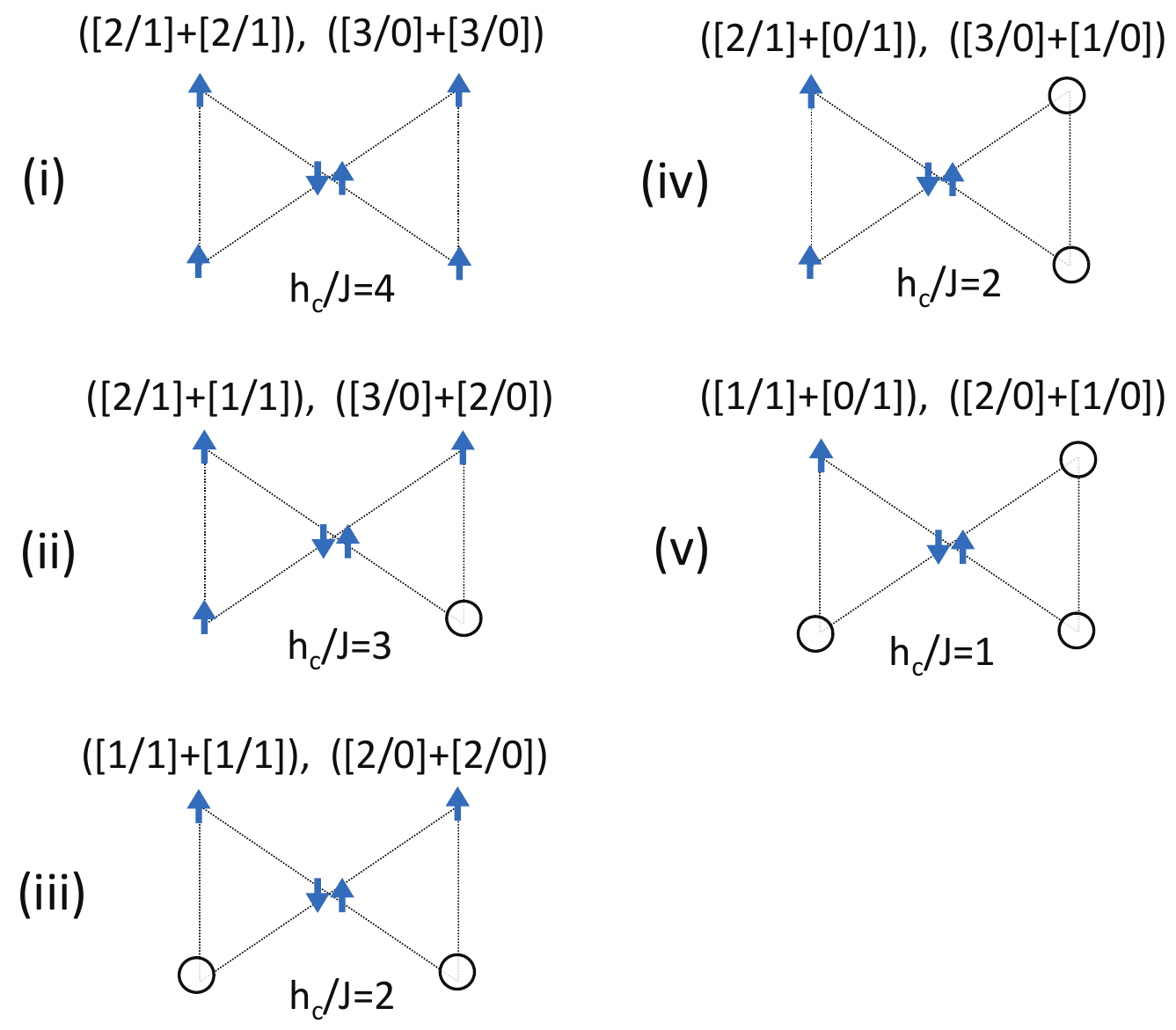}
\caption{
Schematic illustration of spin flip for the kagome lattice.
The up and down spins are denoted by the arrow, 
whereas the deleted spins are denoted by the empty circle.
The crossover values $h_c$ are given there.
}
\label{fig:kag_dilution_flip}
\end{center}
\end{figure}

In the case of the kagome lattice, we consider the energy of 
two corner-sharing triangles. 
The flipping process is illustrated 
schematically in Fig.~\ref{fig:kag_dilution_flip}. 
Deleted spins are denoted by empty circle.
The case where the two triangles have $n=3$ is shown in panel (i). 
When the corner-sharing spin is turned from "down" to "up", 
the configuration changes from [2/1] to [3/0]. 
From Table \ref{kag_configuration}, the crossover magnetic field 
is calculated by
\begin{eqnarray*}
 &~& -J-(h/2) = 3J-3(h/2). 
\end{eqnarray*}
We then obtain $h_c/J=4$.
If one spin is deleted as shown in panel (ii), one triangle has 
$n=3$ and the other triangle has $n=2$. 
When the corner-sharing spin is turned from "down" to "up", 
the configuration changes from ([2/1]+[1/1]) to 
([3/0]+[2/0]). 
From Table \ref{kag_configuration}, the crossover magnetic field 
is calculated by
\begin{eqnarray*}
 &~&(-J-(h/2))+(-J) \\
 &~&\hspace{1cm} =(3J-3(h/2))+(J-2(h/2)). 
\end{eqnarray*}
We then obtain $h_c/J=3$.  
Two cases arise when two spins are deleted: 
(2[1/1]) changes to (2[2/0]), as shown in panel (iii), 
and ([2/1]+[0/1]) changes to ([3/0]+[1/0]), as shown in panel (iv). 
In both cases, the crossover magnetic field is given by $h_c/J=2$.
The case where three spins are deleted is shown in panel (v). 
The corresponding crossover values $h_c$ are obtained 
in the same way as before, and are given 
in Fig.~\ref{fig:kag_dilution_flip}.

This analysis explains the change in the 
configurations given in Fig.~\ref{fig:kag_dilution_config}. 
The origin of the five magnetization plateaus 
in Fig.~\ref{fig:kag_dilution_mag} is thus clearly elucidated, 
as for the triangular lattice.

\subsection{Comparison with "kagome-ice"}

\begin{figure}[t]
\begin{center}
\includegraphics[width=8.4cm]{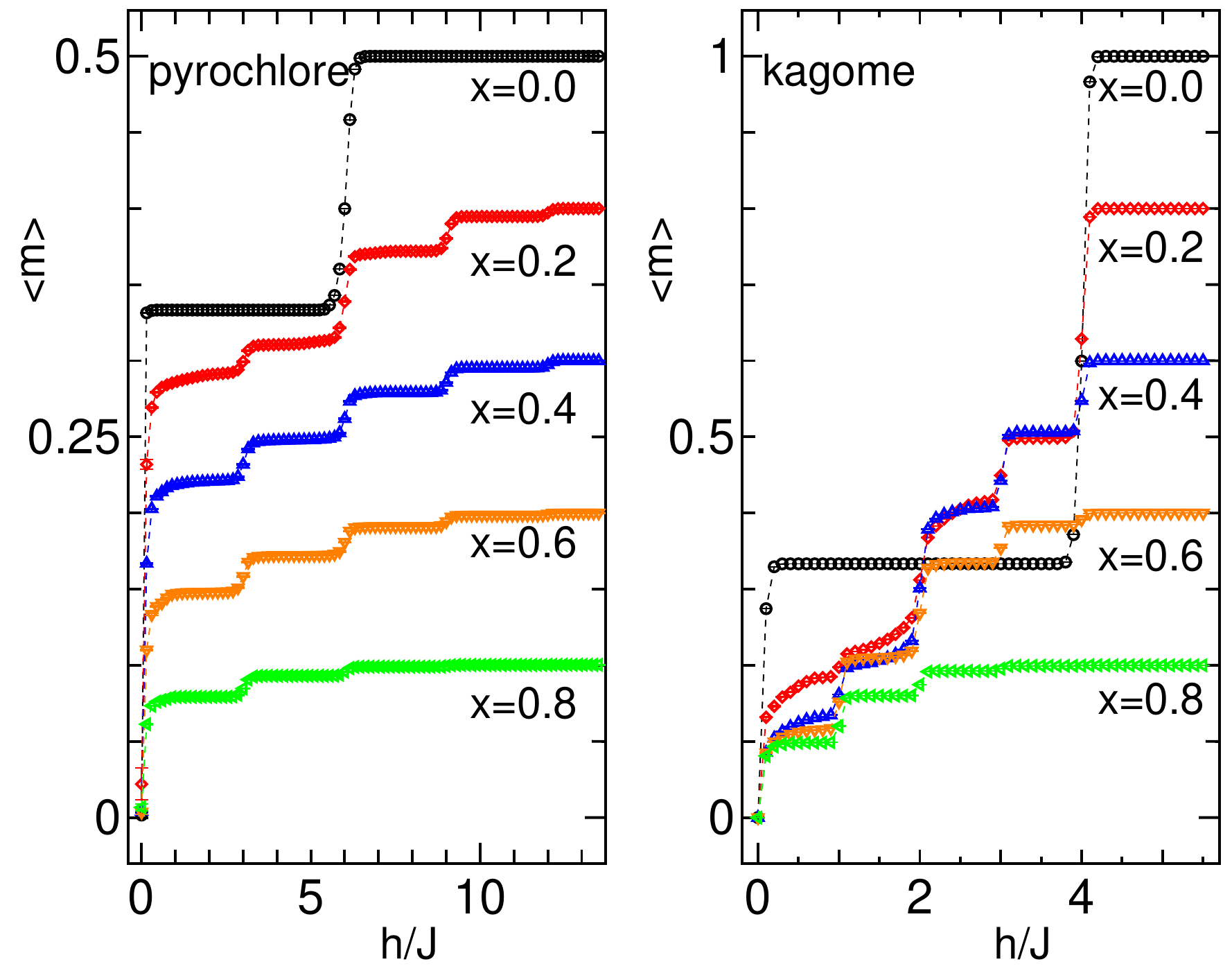}
\caption{
The comparison of the  magnetization curves for the diluted AFM Ising model 
on the pyrochlore lattice \cite{Peretyatko} and on the kagome lattice 
(the present study). 
In the case of pyrochlore lattice the magnetization is applied 
in the [111] direction. 
}
\label{fig:comparison}
\end{center}
\end{figure}

The pyrochlore lattice can be regarded as consisting of alternating kagome 
and triangular layers that become effectively decoupled by a magnetic 
field in the [111] direction. 
Since the spins in the triangular layers are fixed, 
as explained in Sec.~I, Introduction, 
the behavior of the spins in the kagome layers 
has attracted significant interest.
This problem is sometimes referred to as the "kagome-ice" problem 
\cite{Hiroi,Sakakibara}. 
It is interesting to compare the behavior of the "kagome-ice" 
with that of the AFM Ising model on a 2D kagome lattice. 

Figure~\ref{fig:comparison} compares the magnetization curves 
of the diluted AFM Ising models on the pyrochlore lattice \cite{Peretyatko}
and on the kagome lattice (the present study). 
In the case of the "kagome-ice" state for the pure model ($x=0$), 
the magnetization saturates at $m=1/2$, with an intermediate plateau 
at $m=1/3$. It comes from a complex structure of the spin-ice model 
on the pyrochlore lattice.  When a magnetic field is applied 
in the [111] direction, the spins on the [111] positions 
are fixed as 1, but other spins take either $+(1/3)$ or $-(1/3)$ 
\cite{Peretyatko}. 
The magnetization of the kagome lattice saturates at $m=1$ 
for the pure case ($x=0$), with an intermediate plateau 
at $m=1/3$. 

The diluted model of the "kagome-ice" state displays new steps at 
different magnetic fields relative to the pure model  ($h/J=6$), 
namely at $h/J=$ 3, 9, and 12. 
In the case of the kagome lattice, new steps appear 
at $h/J$=1, 2, and 3; the pure case shows a single step 
at $h/J=4$. 
Although there are five plateaus in the magnetization curves 
of the diluted models in the case of both the pyrochlore and 
kagome lattices, the positions of the new transitions are different. 
The plateaus result from the competition 
between the exchange and Zeeman energies, 
which differ in the pyrochlore 
(Table I of Ref.~\cite{Peretyatko}) 
and kagome lattices (Table \ref{kag_configuration} of the present study). 
There are therefore both similarities and dissimilarities 
between the magnetization curves of the diluted model 
for the "kagome-ice" state and for the kagome lattice.

\section{Summary and Discussions}

To summarize, we studied diluted AFM Ising models 
on triangular and kagome lattices 
in a magnetic field using the replica-exchange 
Monte Carlo method for both temperature and magnetic field. 
We observed {\it seven} and {\it five} plateaus 
in the resulting magnetization curves for 
triangular and kagome lattices, respectively. 
These results contrast with the case of the pure model, which displays 
only two plateaus.  The spin configuration within triangles
was investigated, which clearly accounted for the origin 
of multiple magnetization plateaus in the diluted models. 
The present results were compared with those of the diluted 
AFM Ising model on the 3D pyrochlore lattice 
in a magnetic field along the [111] direction, 
a scenario sometimes referred to as the "kagome-ice" problem. 
We discussed the similarity and dissimilarity between 
the magnetization curves for the "kagome-ice" state and 
for the 2D kagome lattice.  Both models have five magnetization 
platueaus, but the positions of magnetization steps are 
different.  It is because the condition of the competition 
between the exchange and Zeeman energies are different 
in both models.

These theoretical results highlight the rich variety 
of effects expected to result from the interplay of 
dilution and magnetic field in frustrated systems. 
Spin-ice materials on the pyrochlore lattice 
have attracted particular attention, and multiple 
magnetization plateaus in the diluted model 
were theoretically proposed quite recently~\cite{Peretyatko}.
The present study on the triangular and kagome 
lattices revealed that this phenomenon 
is general in diluted frustrated systems. 
Future experimental studies will be needed to demonstrate 
them in natural and artificial spin-ice materials \cite{Qi}.

\section*{Acknowledgment}

We thank Vitalii Kapitan, Yuriy Shevchenko, and Hiromi Otsuka 
for valuable discussions. 
The computer cluster of Far Eastern Federal University 
and the equipment of Shared Resource Center 
"Far Eastern Computing Resource" IACP FEB RAS 
were used for computation. 
This work was supported by a Grant-in-Aid for Scientific Research 
from the Japan Society for the Promotion of Science,  
Grants No. JP25400406 and No. JP16K05480.


\begin{thebibliography}{99}

\bibitem{Harris}
 M. J. Harris, S. T. Bramwell, D. F. McMorrow, T. Zeiske, 
 and K.W. Godfrey, Phys. Rev. Lett. {\bf 79}, 2554 (1997).

\bibitem{Ramirez}
 A. P. Ramirez, A. Hayashi, R. J. Cava, R. Siddharthan, and 
 B. S. Shastry, Nature (London) {\bf 399}, 333 (1999).

\bibitem{Bramwell}
 S. T. Bramwell and M. J. P. Gingras, Science {\bf 294}, 1495 (2001).

\bibitem{Pauling}
 L. Pauling, J. Am. Chem. Soc. {\bf 57}, 2680 (1935).

\bibitem{Harris98}
 M. J. Harris, S. T. Bramwell, P. C. W. Holdsworth, 
 and J. D. M. Champion,
 Phys. Rev. Lett. {\bf 81}, 4496 (1998).

\bibitem{Moessner}
 R. Moessner and S. L. Sondhi, Phys. Rev. B {\bf 68}, 064411 (2003).

\bibitem{Isakov}
 S. V. Isakov, K. S. Raman, R. Moessner, and S. L. Sondhi, 
 Phys. Rev. B {\bf 70}, 104418 (2004).

\bibitem{Matsuhira02}
 K. Matsuhira, Z. Hiroi, T. Tayama, S. Takagi, and T. Sakakibara,
 J. Phys.: Condens. Matter {\bf 14}, L559 (2002).

\bibitem{Hiroi}
 Z. Hiroi, K. Matsuhira, S. Takagi, T. Tayama, and T. Sakakibara,
 J. Phys. Soc. Jpn. {\bf 72}, 411 (2003).

\bibitem{Sakakibara}
 T. Sakakibara, T. Tayama, Z. Hiroi, K. Matsuhira, and S. Takagi,
 Phys. Rev. Lett. {\bf 90}, 207205 (2003).

\bibitem{Higashinaka}
 R. Higashinaka, H. Fukazawa, and Y. Maeno, 
 Phys. Rev. B {\bf 68}, 014415 (2003).

\bibitem{Fukazawa}
 H. Fukazawa, R. G. Melko, R. Higashinaka, Y. Maeno, 
 and M. J. P. Gingras, 
 Phys. Rev. B {\bf 65}, 054410 (2002).

\bibitem{Ke}
 X. Ke, R. S. Freitas, B. G. Ueland, G. C. Lau, M. L. Dahlberg, 
 R. J. Cava, R. Moessner, and P. Schiffer,
 Phys. Rev. Lett. {\bf 99}, 137203 (2007).
\bibitem{Lin}
 T. Lin, X. Ke, M. Thesberg, P. Schiffer, R. G. Melko, and M. J. P. Gingras,
 Phys. Rev. B {\bf 90}, 214433 (2014).
\bibitem{Scharffe}
 S. Scharffe, O. Breunig, V. Cho, P. Laschitzky, M. Valldor,
 J. F. Welter, and T. Lorenz,
 Phys. Rev. B {\bf 92}, 180405(R) (2015).
\bibitem{Shevchenko}
 Y. Shevchenko, K. Nefedev, and Y. Okabe, 
 accepted for publication in Phys. Rev. E,
 arXiv:1705.05446 (2017).

\bibitem{Peretyatko}
 A. Peretyatko, K. Nefedev, and Y. Okabe, 
 Phys. Rev. B {\bf 95}, 144410 (2017).

\bibitem{Hukushima}
  K. Hukushima and K. Nemoto, J. Phys. Soc. Jpn. {\bf 65}, 1604 (1996).

\bibitem{Yao} 
 X. Yao, 
 Solid State Commun. {\bf 150}, 160 (2010).

\bibitem{Zukovic} 
 M. \u{Z}ukovi\u{c}, M. Borovsk\'y, and A. Bob\'ak,
 Phys. Lett. A {\bf 374}, 4260 (2010).

\bibitem{Borovsky} 
 M. Borovsk\'y, M. \u{Z}ukovi\u{c}, and A. Bob\'ak,
 Acta Physica Polonica A {\bf 126}, 16 (2014).

\bibitem{Qi} 
 Y. Qi, T. Brintlinger, and J. Cumings,
 Phys. Rev. B {\bf 77}, 094418 (2008).

\end{thebibliography}
\end{document}